\documentclass[12pt,preprint]{aastex}
\usepackage{amsmath,amssymb,graphicx,natbib}

\newsavebox{\astrutbox}
\sbox{\astrutbox}{\rule[-5pt]{0pt}{20pt}}

\newcommand{\gapprox}{\lower.4ex\hbox{$\;\buildrel >\over{\scriptstyle\sim}\;$}}
\newcommand{\lapprox}{\lower.4ex\hbox{$\;\buildrel <\over{\scriptstyle\sim}\;$}}

\bibpunct[; ] {(}{)}{,}{}{}{,}
\slugcomment{To be submitted to ApJ.} 
\shorttitle{GLOBAL PROPERTIES OF CONVECTIVE DISKS} 
\shortauthors{Bodo, Cattaneo, Mignone, Ponzo \& Rossi} 

\begin{document} 
 
\title{GLOBAL PROPERTIES OF FULLY CONVECTIVE ACCRETION DISKS FROM LOCAL SIMULATIONS} 
 
 \author{G. Bodo\altaffilmark{1},        
         F. Cattaneo\altaffilmark{2,3}, 
         A. Mignone\altaffilmark{4},
         F. Ponzo\altaffilmark{1,4},
         P. Rossi\altaffilmark{1} }
  
 \altaffiltext{1}{INAF, Osservatorio Astrofisico di Torino, Strada Osservatorio 20, 10025 Pino Torinese, Italy}
 
 \altaffiltext{2}{The Computation Institute, The University of Chicago, 
              5735 S. Ellis avenue, Chicago IL 60637, USA}

 \altaffiltext{3}{Department of Astronomy and Astrophysics, The University of Chicago, 
              5640 S. Ellis avenue, Chicago IL 60637, USA}

 \altaffiltext{4}{Dipartimento di Fisica Generale, Univesit\'a di Torino, via Pietro Giuria 1, 10125 Torino, Italy} 
 
\begin{abstract} 
We present an approach to deriving global properties of accretion disks from the knowledge of local solutions derived from numerical simulations based on the shearing box approximation. The approach consists of a two-step procedure. First a local solution valid for all values of the disk height is constructed by piecing together an interior solution obtained numerically with an analytical exterior radiative solution. The matching is obtained by assuming hydrostatic balance and radiative equilibrium. Although in principle the procedure can be carried out in general, it simplifies considerably when the interior solution is fully convective. In these cases, the construction is analogous to the derivation of the Hayashi tracks for protostars. The second step consists of piecing together the local solutions at different radii to obtain a global solution. Here we use the symmetry of the solutions with respect to the defining dimensionless numbers--in a way similar to the use of homology relations in stellar structure theory--to obtain the scaling properties of the various disk quantities with radius.
\end{abstract} 

\section{Introduction}
Determining the structure of accretion disks remains one of the outstanding problems in astrophysics. Much of what is known theoretically is based on simulations defined within the framework of the shearing box approximation \citep{Hawley95}. Because of the local nature of this approximation some further assumptions are needed to extract from the local simulations global disk properties like the radial dependence of the various disk properties and, ultimately, the accretion rate. One way  to proceed, for instance, is to assume that the disk is an $\alpha$-disk \'a la  \citet{SS73} and then extract the value of $\alpha$ from the shearing box simulations. Although this procedure is perfectly reasonable, it is a little bit wasteful, in the sense that all the information from the numerical simulations are combined into a single number. Also, there are situations  in which defining an $\alpha$ parameter unambiguously from the simulations is not entirely straightforward \citep{Bodo15}.
All in all, one feels that in spite of everything, shearing box simulations tell us quite a lot about the disk structure, and that it 
should be possible to piece together the results of shearing box simulations in such a way as to construct an object that looks locally like a shearing box and globally like an accretion disk. 

In the present paper we examine the feasibility of effecting such a construction, and argue that under suitable conditions, namely when the disk is fully convective, the construction can actually be carried out quite straightforwardly. When discussing fully or partially convective disks, it is important to be clear about the role of the convection within the overall disk dynamics. One possibility is to assume that the turbulence that gives rise to the enhanced angular momentum transport is convective in origin. This being the case, the convection carries not only thermal energy upwards, but also angular momentum outwards. In other words, convection and disk turbulence are one and the same, and the angular momentum is transported mostly by the $r-\theta$ component of the Reynolds stresses. It is now believed that this is probably never the case \citep{Cabot96, Stone96}. More likely, is that the origin of the disk turbulence is not convective but probably associated with magnetically mediated instabilities, and that the role of the convection is just to carry the heat vertically upwards when radiation fails to do the job. This second possibility has been confirmed by  several numerical experiments showing that if the heat generated by turbulence driven by the Magneto-Rotational-Instability (MRI) cannot make it to the disk surface by radiative transport, convection sets in. In these simulations, the outwards transport of angular momentum is mostly associated with magnetic stresses; the role of the convection is mainly to contribute to the vertical energy transport, and to give rise to more efficient  dynamo processes \citep{Bodo12, Bodo13, Bodo15, Hirose14, Hirose15}.

As we shall see presently, the construction of a global solution is best described in terms of a two-step procedure. The first step is to match together an interior and an exterior solution so that certain continuity conditions are satisfied across the interface. The interior solution is obtained numerically from shearing box simulations,  the exterior solution is  obtained by assuming radiative equilibrium and hydrostatic balance. Although this procedure could be carried out in general, it simplifies considerably if the interior solution is fully convective. The reason for the simplification is that fully convective solutions are independent of the detailed form of the thermal conduction (opacity) and therefore the solutions depend on fewer parameters than in the general case. We note that this matching procedure is analogous to that used to construct the Hayashi tracks for fully convective, pre-main sequence stars \citep{Hayashi61}. 

In the second step the scaling relations between the disk radius and the dimensionless numbers that define the local solutions are used to construct a global solution that is valid for all radii. This is similar to the use of homologous relationships in stellar structure theory to determine various laws between the stellar mass, luminosity, effective temperature, etc. \citep[see, for instance,][]{Prialnik11}. Here, the basic assumption, is that the disk remains close to  Keplerian, and that therefore there is a specific, known relationship between the radius and the orbital period. This is an  assumption that amounts to stating that the turbulence within the disk, whatever its origin, saturates by dissipative losses and not by modifying the basic rotational profile. Finally we argue that this global solution can, at least in principle, be used to compute the mass accretion rate once the mass of the central object and the disk's mass within a given radius are specified.

\section{The vertical structure}
 \label{vert_struct} 
Our aim is to construct the consistent vertical structure of an accretion disk at a given radial position using as a starting point the  results of our numerical  simulations of shearing boxes with vertical heat transport  \citep{Bodo12, Bodo13, Bodo15}. Therein, it is shown that the nondimensional shearing box equations with radiative boundary conditions depend only on two dimensionless parameters. These can be chosen to be the P\'eclet number, $Pe$, which measures thermal diffusivity, and the radiation parameter, $\Sigma_r$, 
 which measures the efficiency of the black-body radiating boundary \citep{Bodo15}. 
Depending on the value of the P\'eclet  number,  two families of stationary solutions are observed, one at low values of $Pe$, in which heat is carried by conduction and one, at high values of $Pe$, in which it is carried by convection. In this second case the solutions become asymptotically independent of $Pe$,  and thus depend only on the single parameter $\Sigma_r$. This considerably simplifies the analysis; thus from now on we concentrate on this latter case. 

\begin{figure}[htbp]
   \centering
      \includegraphics[width=5cm]{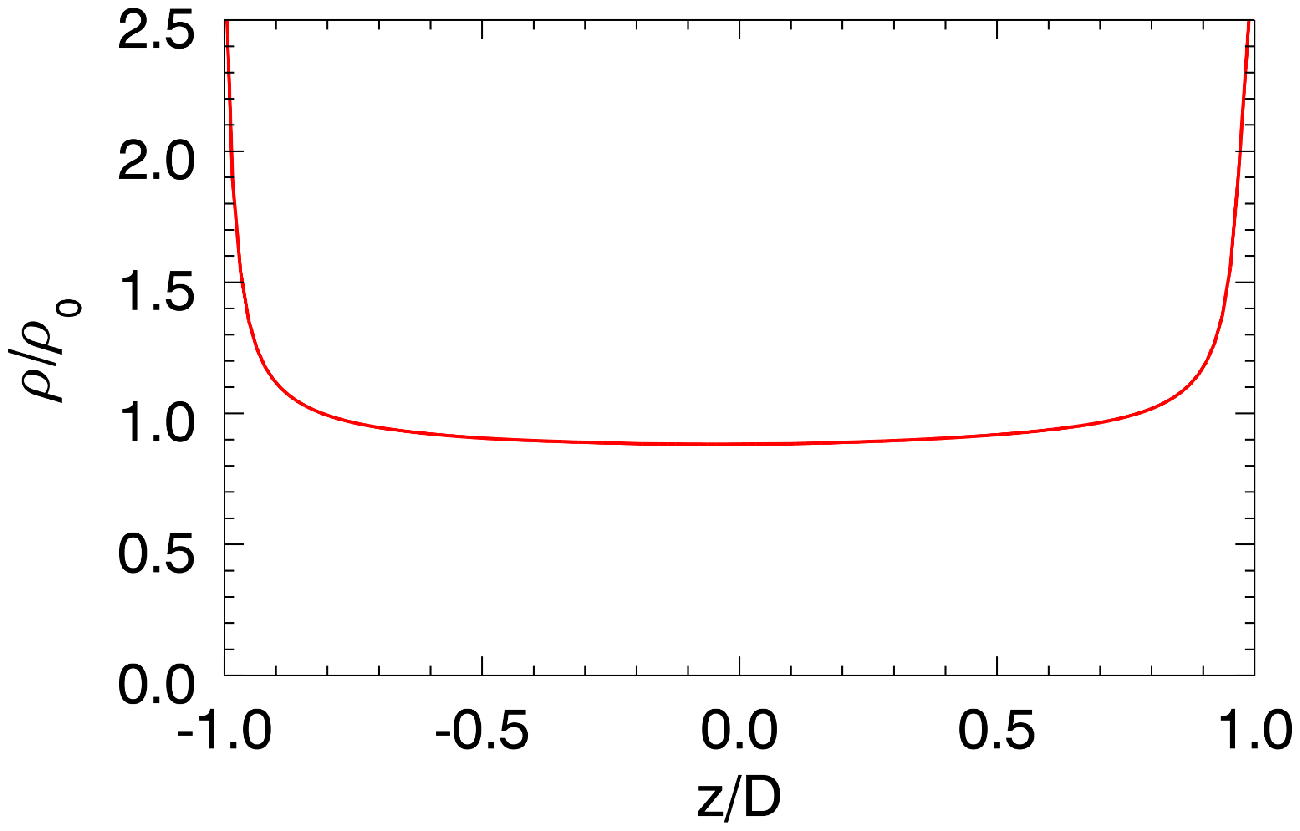} 
      \includegraphics[width=5cm]{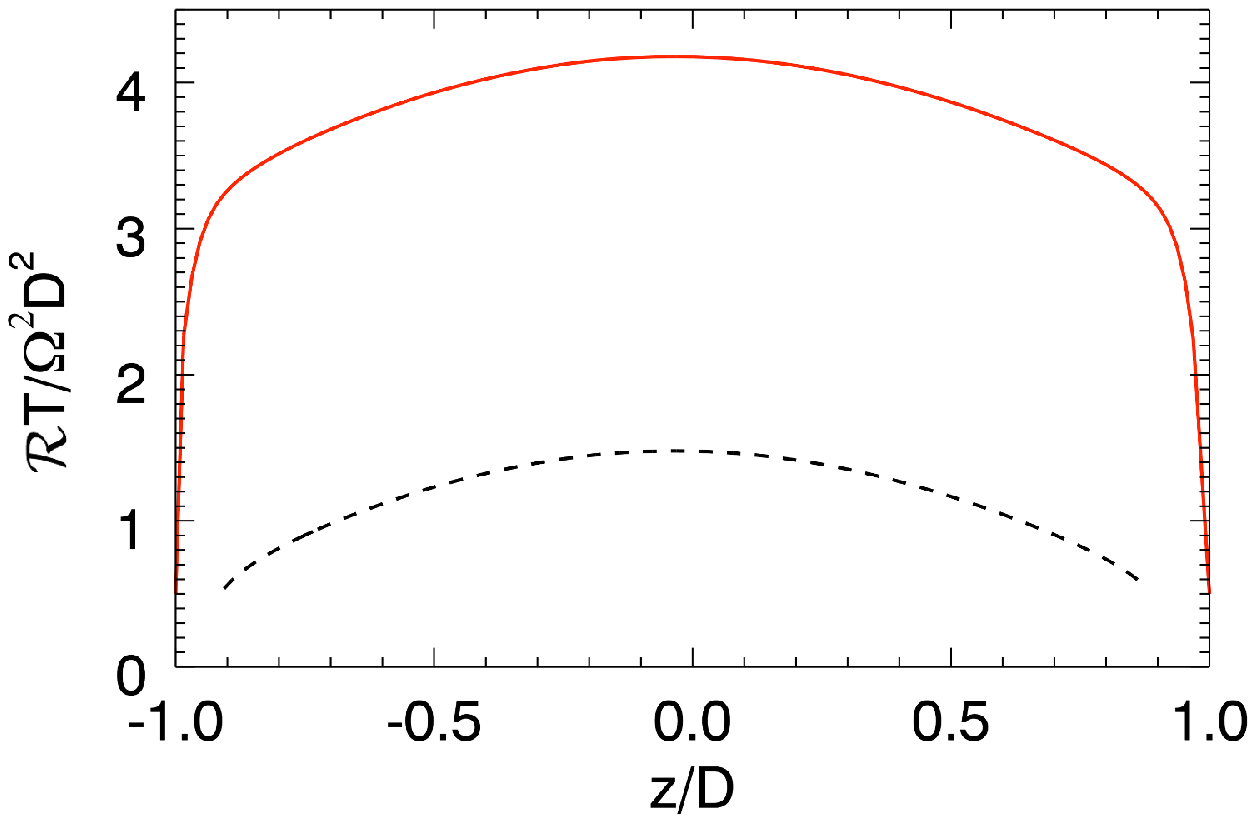} 
      \includegraphics[width=5cm]{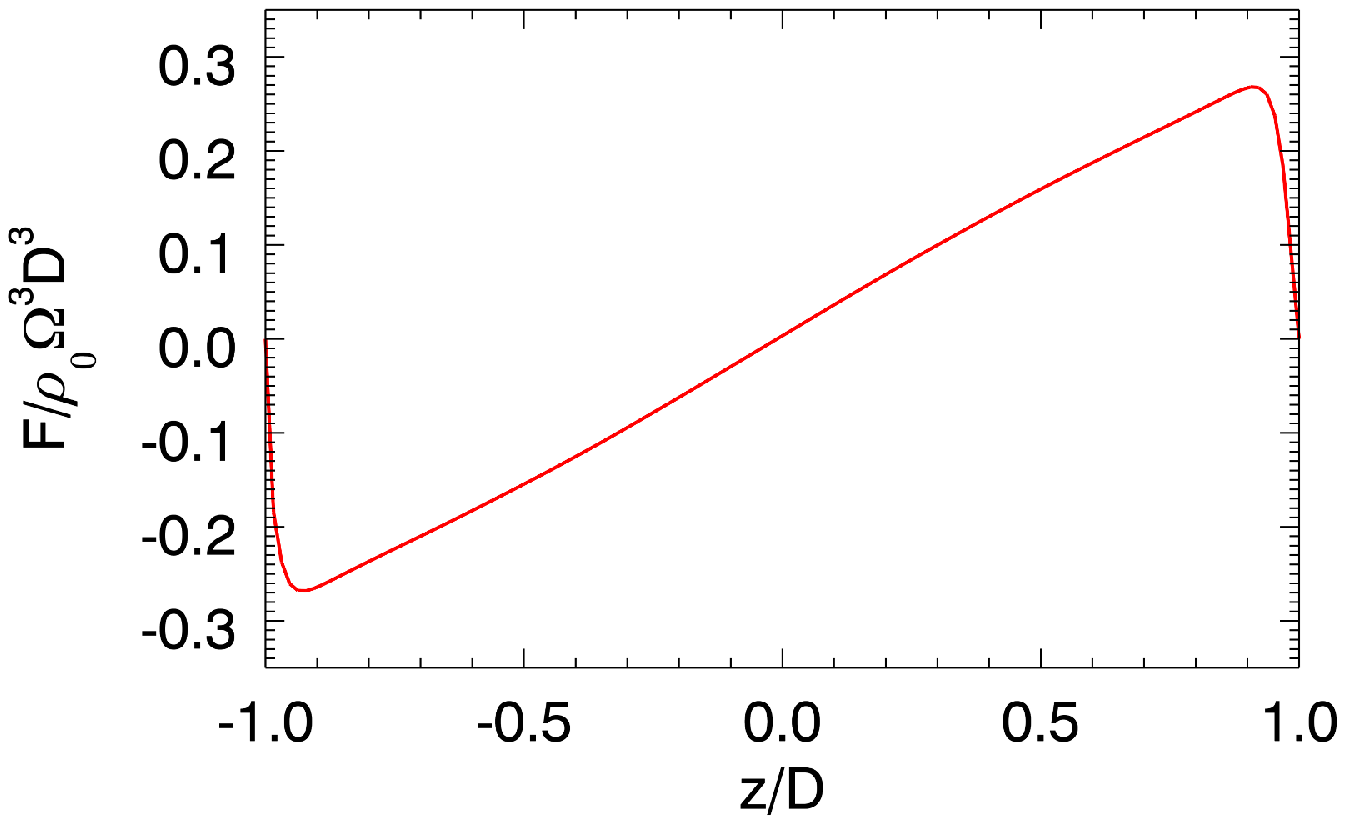} 
   \caption{Horizontal and time averaged density (left panel), temperature (midle panel) and convective flux (right panel) as a function of the vertical coordinate, for a  shearing box simulation with $Pe = 1,000$ and $\Sigma_r=20.0$. The domain size was $8\times6\times2$ with a resolution of $512\times384\times128$. For a full discussion of the numerical results, see \citet{Bodo15}. In the middle panel, the lower curve is the one used for the matching. }
   \label{fig:convective_solutions}
\end{figure}

The construction of a solution that is valid for all heights requires the matching of an interior solution--obtained numerically, with an exterior solution--obtained analytically.   The former is best expressed in terms of {\it dimensionless} quantities, the latter in terms of {\it dimensional} ones. Thus, it becomes important to be able to swap between dimensional and dimensionless quantities and to keep track of which quantities are which. In what follows all quantities are assumed to be {\it dimensional} unless otherwise stated, and we adopt the following units: the average mass density, $\rho_0$, as the unit of density, half the layer thickness (shearing box height), $D$, as the unit of length, and the rotational frequency $\Omega$ as the unit of frequency. Thus, for instance, the heat flux is measured in units of $\rho_0 D^3 \Omega^3$ while the temperature is measured in units of $D^2 \Omega^2 /{\cal R}$, where ${\cal R}$ is the gas constant. Furthermore, with these choices, $Pe$ and $\Sigma_r$ are defined by
\begin{equation}
Pe=\frac{D^2 \Omega}{\kappa}, \qquad \Sigma_r=\frac{\sigma \Omega^5D^5}{{\cal R}^4 \rho_0},
\label{eq:definitions}
\end{equation}
where $\kappa$ is a characteristic value of the thermal diffusivity and $\sigma$ is Stefan-Boltzmann constant.
A representative example  of  an interior solution is illustrated in Fig. \ref{fig:convective_solutions} which shows the average vertical profiles of density, temperature and heat flux, for a typical case in the convective regime.

The matching procedure can be carried out directly with the numerical solutions, however, as a 
first approximation and for simplicity we model the numerical solutions  by profiles that are constant for the density, $\rho$, parabolic for the temperature, $T$, and linear for the heat flux, $F$. These can be written as 
\begin{equation}
\rho(z) = \rho_0 \;,
\label{eq:rho_conv}
\end{equation}
\begin{equation}
T(z) = T_c  - A z^2 \;,
\label{eq:T_conv}
\end{equation}
\begin{equation}
F(z) =  H(\Sigma_r) z \;.
\label{eq:F_conv}
\end{equation}
The constant $A$ is fixed by assuming hydrostatic balance which gives $A=1/2~ \Omega^2$. 
Here, $T_c$ corresponds to the equatorial temperature; it must be chosen so that the temperature profile matches the ``effective'' temperature at the upper boundary (lower curve in Figure \ref{fig:convective_solutions}). This is an important point that will be further discussed later. 
The quantity $H$ in equation (\ref{eq:F_conv}) represents the average energy dissipation rate per unit volume whose dimensionless value is obtained from the simulations and observed to be approximately constant--i.e. $z$ independent, hence the linear profile for $F$. 

We assume that the top of the convective region corresponds to the disk photosphere on top of which we  have a  radiative region of optical depth unity whose thickness is assumed to be much smaller than $D$.
A relation connecting the height $D$ of the convective region, which in our assumptions corresponds to the height of the disk, to its average density can be obtained by joining the fully convective interior to the radiative photosphere following a procedure similar to that used to construct the  Hayashi tracks on the H-R diagram    \citep[see e.g.][]{Prialnik11}. 

First of all we impose, in the radiative photosphere,  the condition of hydrostatic equilibrium
\begin{equation}
\frac{dP}{dz} = -\Omega^2 D \rho \;,
\end{equation}
where we assume that the thickness of the radiative layer is small and therefore we neglect the variation of gravity with height. Integrating from infinity, where we assume that the pressure vanishes, to the matching point we obtain
\begin{equation}
P_D = \Omega^2 D \int^\infty_{D} \rho dz \;,
\label{eq:hyd_eq}
\end{equation}
where $P_D$ is the pressure at the matching point $z = D$. The optical thickness  of the radiative part must be unity and therefore
\begin{equation}
\int^\infty_{D}  \varkappa \rho dz =  \overline \varkappa  \int^\infty_{D} \rho dz = 1
\label{eq:taum}
\end{equation}
where $\varkappa$ is the opacity and $\overline \varkappa $ is its average over the radiative region. We can further  approximate 
Eq.~(\ref{eq:taum}) by taking $\overline \varkappa $ to be the opacity at $z = D$ and expressing it as a power law in density and temperature. The density at the matching point is equal to $\rho_0$ and the temperature is equal to the effective temperature $T_{eff}$, therefore
\begin{equation}
\overline \varkappa = \varkappa_0 \rho_0^a T_{eff}^b
\label{eq:kappa}
\end{equation}
 and
\begin{equation}
 \varkappa_0 \rho_0^a T_{eff}^b   \int^\infty_{D} \rho dz  = 1\;.
\end{equation}
Thus Eq.~(\ref{eq:hyd_eq}) can then be rewritten as 
\begin{equation}
P_D = \Omega^2 D \frac{1}{\varkappa_0 \rho_0^a T_{eff}^b} \;.
\end{equation}
An additional relation between density, pressure and temperature at the matching point is provided by the perfect gas equation of state
\begin{equation}
P_D = \frac{\cal R}{\mu} \rho_0 T_{eff}
\end{equation}
where $\mu$ is the mean molecular weight. This finally gives
\begin{equation}
\frac{\cal R}{\mu} \rho_0 T_{eff} = \Omega^2 D  \frac{1}{\varkappa_0 \rho_0^a T_{eff}^b} \;.
\label{eq:matching1}
\end{equation}
Furthermore, $T_{eff}$ can be related to the heat flux at the top of the convective region by the relation
 \begin{equation}
\sigma  T^4_{eff}  = {\cal H}(\Sigma_r) \rho_0 \Omega^3 D^3\;,
\label{eq:matching2}
\end{equation}
where  ${\cal H}(\Sigma_r)$ is dimensionless. 

\begin{figure}[htbp]
   \centering
      \includegraphics[width=10cm]{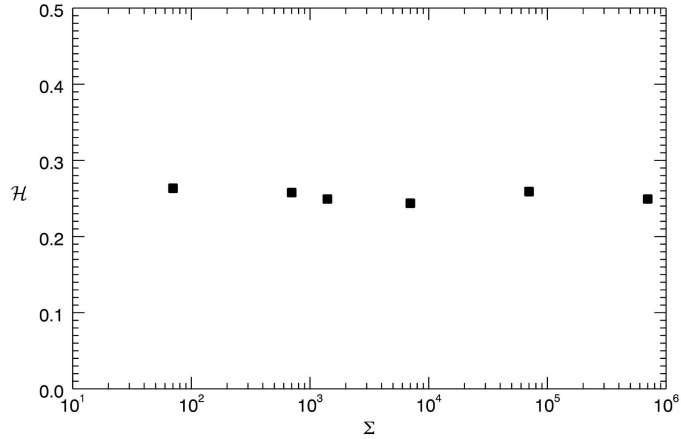} 
   \caption{Plot of $\cal H$ as a function of $\Sigma_r$ for a series of shearing box simulations with $Pe=$ 1,000. As can be seen, the value of $\cal H$ is nearly constant over the range. Details of the numerical work can be found in \citet{Ponzo15}. }
   \label{fig:sigma}
\end{figure}

Eqs. (\ref{eq:matching1}) and (\ref{eq:matching2}) represents, for a given $\Omega$,  two equations for the three unknown $\rho_0$, $T_{eff}$ and $D$. Thus, every quantity  can, in principle, be expressed as a function of $\rho_0$. To do this, however,  the functional form of ${\cal H}(\Sigma_r)$ is needed. To obtain it, we performed a series of shearing box simulations with values of $\Sigma_r$ spanning four order of magnitudes, ranging from $10^{2}$ to $10^6$ and found that ${\cal H}(\Sigma_r)$ is nearly independent of $\Sigma_r$ over this range (see Fig.~ \ref{fig:sigma}). Consequently we assume it to be a constant.
The expression for the height of the disk then becomes
\begin{equation}
  D = \left( \frac{\cal H}{\sigma} \right)^{-(1 + b) / (3 b - 1)} \left(
  \frac{{\mu}}{{\cal R}  {\varkappa}_0 } \right)^{4 / (3 b - 1)} \Omega^{(5 -
  3 b) / (3 b - 1)} ~\rho_0^{- (5 + 4 a + b) /( 3 b - 1)}  \;.
  \label{eq:discD} 
\end{equation}
Finally, Eq.~(\ref{eq:T_conv}) can be used to obtain the value of the central temperature. In this way we have fully specified the vertical disk structure for given values of $\Omega$ and $\rho_0$. In the next section we patch together these solutions as a function of the radial distance from the central object to obtain the global disk structure.

\section{The global disk structure}
 \label{global_struct} 
In Eq.~(\ref{eq:discD}) the disk height is implicitly expressed as a function of the radial distance, $r$ from the central object through $\Omega$ and the density $\rho_0$. 
The required dependence of $\Omega$ on $r$ can be obtained by assuming that the disk is Keplerian, whereby
\begin{equation}
\Omega(r)= \sqrt{GM/r^3},
\label{eq:omega}
\end{equation}
and $G$ and $M$ are the gravitational constant and the mass of the central object, respectively. 
The radial dependence of $\rho_0$ can be obtained from the condition that the mass accretion rate, $\dot M$, in a steady state, must be independent of the radius. The mass accretion rate,  in turn, can be derived from the angular momentum conservation \citep[see e.g.][]{Pringle81} that, at large distances, can be expressed as 
\begin{equation}
\dot M \Omega r^2 = 4 \pi S r^2 D \;.
\label{eq:mdot}
\end{equation}
Here,  $S$ is the value of the $r\phi$ component of the turbulent stress tensor given by
\begin{equation}
S(\Sigma_r) = {\cal S}(\Sigma_r) \rho_0 \Omega^2 D^2 \;,
\label{eq:stresses}
\end{equation}
where ${\cal S}(\Sigma_r)$ is its nondimensional value obtained from the numerical simulations. We note here that, up to normalization factors of order unity,  ${\cal S}$ is the same as the $\alpha_{SS}$ parameter in  \citet{SS73}. Equations (\ref{eq:mdot}) and (\ref{eq:stresses}) can be combined to express the density $\rho_0$ as
\begin{equation}
\rho_0 = \frac{\dot M}{4 \pi {\cal S}(\Sigma_r) D^3 \Omega }.
\label{eq:rho0}
\end{equation}
Again we note that in general ${\cal S}$ is a function of $\Sigma_r$, however in the shearing box formulation ${\cal S}$ and ${\cal H}$ are proportionally related  \citep{Hawley95}, in fact in a steady state ${\cal S} = 2/3 {\cal H}$. Therefore, since we have argued that ${\cal H}$ is approximately constant, so is ${\cal S}$. Indeed, the values in  Fig. \ref{fig:sigma} correspond to a value of the $\alpha_{SS}$ parameter of approximately 0.16 normalized to the disk half thickness $D$.

Finally inserting Eq.~(\ref{eq:rho0}) in Eq.~(\ref{eq:discD})   we express $D$ as a function only of $\Omega$ and, therefore, for a given mass of the central object, as a function of the radial distance and of the mass accretion rate:
\begin{equation}
D = \left( \frac{1}{4 \pi }\right)^{(5+4a+b)/4 \beta}
\left(  \frac{1}{\cal S} \right)^{(1+a)/\beta}
\left(\frac{3}{2 \sigma}  \right)^ {(1+b)/4 \beta}
\left(  \frac{\mu}{{\cal R} \varkappa_0} \right)^{-1/\beta}
\dot M^{(5+4a+b)/4 \beta} \quad
\Omega^{-(5+2a-b)/2 \beta}  ,
\label{eq:height}
\end{equation}
where $\beta=(4 + 3a)$ and we have used the steady state relation ${\cal S} = 2/3 {\cal H}$. For comparison, we note that this expression is equivalent to equation (2.16) or (2.19) in \citet{SS73} that give the disk height: $z_0$ in their notation. The effective temperature can be  obtained from Eqs.~(\ref{eq:matching2}) and (\ref{eq:rho0}) as
\begin{equation}
T_{eff} = \left( \frac {3}{8 \pi \sigma } \right)^{1/4} \dot M^{1/4} ~~\Omega^{1/2}.
\end{equation}

\section{Discussion and Conclusion} \label{conclusions}
We have outlined a procedure to construct a global model of an accretion disk by piecing together the numerical solutions based on the shearing box approximation. The method naturally follows in two steps, one in which the local numerical solutions are matched to an exterior radiative atmosphere in hydrostatic balance, another in which the scaling relations are used to extend the local construction to the entire disc. Although the methods can be applied in general, it simplifies considerably if the solution in the disk interior is fully convective. Indeed, in the present paper we have restricted our analysis to such cases. The two steps, matching to a radiative atmosphere, and using scaling relations are similar to the construction in stellar structure theory of the Hayashi tracks for fully convective proto-stars \citep{Hayashi61}, and of the derivation of the mass-luminosity relations for main sequence homologous stars.

The final result gives the complete structure of the disk as a function of two parameters; $M$ and $\dot M$ once the physical properties of the accreting  material, like its equation of state and opacity law are specified. This may seem somewhat paradoxical. One has the feeling that for  infalling material with fixed (known) physical properties, hydrogen, say, the accretion rate should be determined only by the mass of the central object. Why, then, is that not the case  here? To understand the nature of the problem it suffices to examine Eq. (\ref{eq:height}) which specifies the height of the disk as a function of $\Omega$ and hence of the radius. The first three terms on the right hand side are fixed by the solutions of the shearing box problem and by the physical properties of the accreting gas. The last two depend on $\dot M$ and $M$. Thus for a fixed $M$ and fixed physical properties the global solution consists of a one parameter family of discs of different thicknesses--given by $D$, and depending on $\dot M$; raising the question why is the accretion rate not uniquely defined? There appear to be a residual  ambiguity in the absolute vertical scale of the disk. The origin of this ambiguity is related to the decoupling of the length scales for radial and vertical distances. Typically, the unit of length is isotropic and applies equally to radial and vertical coordinates. This is indeed the case in the global disk equations. However, the formulation of the shearing box equations as a local approximation decouples the radial and vertical coordinates. The position of a shearing box within a disk is only encoded by $\Omega$, not by $r$. Thus it is impossible to specify the size or height of a shearing box in terms of the absolute radial coordinate. Our construction of the global solution by piecing together appropriately scaled local solutions produces the correct $r$ dependence but not the absolute scale. This has to be reintroduced separately, by, for example, specifying $\dot M$. Alternatively,  a unique accretion rate can be recovered by specifying the total mass within a cylinder of radius $r$. 

These considerations notwithstanding we have constructed a global model for the disk. It is useful to re-examine what assumptions were made that allowed the construction to proceed and, to what extent, they were reasonable. These come in two varieties: weak assumptions that are general to any type of disk and strong assumptions that are specific to our case.  In order to move from a local to a global model we assumed that the disk was stationary--that allowed the derivation of  a relationship between $\dot M$ and $\rho_0$, also we assumed that the disk's rotational profile was Keplerian to obtain a relationship between $\Omega$ and $r$. Both these are weak assumptions that do not require much in the way of justification. It is reasonable to assume that at least some disks are nearly Keplerian and that they may spend some  of their existence in a quasi-stationary state. Much stronger assumptions are those concerning the locality of the disk structure, the convective nature of the disk interior, the thinness of the transition layer between optically thick and optically thin regions, and the use of numerical solutions obtained with impenetrable boundaries and fixed diffusivity to model a physical layer with no boundaries and strongly varying diffusivity. The first three assumptions are physical, the last is procedural. We discuss them in turn. 

The use of the shearing box approximation to determine the disk structure implicitly assumes that there is no global, or at any rate, long-range organization of the disk. This assumption can only be verified by comparing with full disk simulations.  The interior of a disk will become convective if the P\'eclet number is large; which, in turns, requires that the disk material be very optically thick. In order for the transition layer between convective interior and radiative exterior to be geometrically thin, the opacity must be a strongly decreasing function of temperature. These requirements could be satisfied, possibly, by plasmas near their ionization conditions. For example, depending on the density, hydrogen, has such a transition at temperatures ranging between a few thousand to ten thousand degrees. At higher temperatures  hydrogen becomes fully ionized, most of the contributions to the opacity come from processes involving free electrons and the opacity decreases with increasing temperature. Thus it is not immediately obvious that  both conditions, a strong (decreasing) temperature dependence and large P\'eclet number in the interior can be simultaneously satisfied. However, because we have an explicit solution we can check {\it a posteriori} if that is the case. 

\begin{figure}[htbp]
   \centering
   \includegraphics[width=10cm]{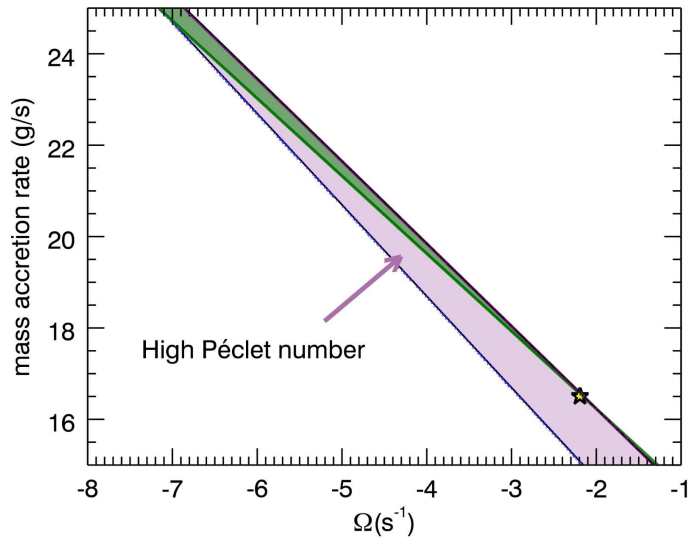} 
   \caption{Regions of model consistency in the $\Omega - {\dot M}$ plane. The pink region corresponds to convective solutions. The green region corresponds to convective solution with thin``transition" regions. Here we have used a value of ${\cal H}=0.26$ from Fig.~\ref{fig:sigma}. For shearing box simulations  ${\cal S} = 2/3 ~{\cal H}$. The star corresponds to one of the convective solutions from \citet{Hirose14}.
}
   \label{fig:consistency}
\end{figure}

We now illustrate how this could be done in practice. In the radiative exterior we assume an opacity law of the form (\ref{eq:kappa}) with $a=0.5$ and $b=9$. These values are not unreasonable for hydrogen at a few thousand degrees and moderate densities \citep{Hansen04}. Then for each value of $\dot M$ and $\Omega$ we construct a solution and evaluate an average P\'eclet number in the interior based on Kramer's opacity law, say, and the scale-height at $D$ of the exterior radiative solution based on the opacity law given above. The results are summarized in Fig.~\ref{fig:consistency}. The region to the left of the purple line correspond to solutions with $Pe >>1$,
the region to the right of the blue line correspond to solutions with $T_{eff} \geq 3000K$, for which our assumed exterior opacity law is justified, and the region to the right of the green line correspond to solutions with scale heights at $D$ much smaller than $D$ itself. Thus solutions in the green region satisfy all the required criteria and are therefore self-consistent.  The pink region corresponds to solutions that are convective but the transition to the radiative state is not sharp. For comparison, the star corresponds to the values in one of the convective solutions recently found by \citet{Hirose14}. Indeed in their solutions the transition region  is not very sharp and, correspondingly, the star is just inside the pink region.   We note that It is important to understand why we have chosen to exclude these solutions from our construction. The numerical basis for our models were calculated by considering a computational domain with impenetrable, stress-free boundaries in the vertical and a constant value of the thermal diffusivity. The latter was chosen to be small to ensure that the P\'eclet number was large. The result are solutions with sharp thermal boundary layers at the boundaries as can be seen in Fig. \ref{fig:convective_solutions}. What is being considered here, instead, are solutions in which the temperature varies smoothly, but the opacity, and hence the thermal diffusivity varies sharply. In other words we have traded a constant $\kappa$, impenetrable boundaries, and large temperature gradients for moderate temperature gradients, no clear cut boundary, and a sharply varying $\kappa$. In general one would not expect the two situations to be interchangeable unless the regions with the large gradients are geometrically thin. The other justification for this procedure is that it should be possible to arrange things so that the trade-off can be carried out in such a way that the total energy flux, and hence, the total energy generation rate is preserved. This is precisely why, in Eq.~(\ref{eq:T_conv}) that describes the temperature profile {\it without} the boundary layers, we chose $T_c$ so that it matched $T_{eff}$ at the boundaries. Thus the two cases, one with  thermal boundary layers and one without do not have the same interior temperature but have the same energy flux, and hence the same accretion rate. 

On the basis of Fig.~\ref{fig:consistency} we conclude that it is probably unlikely that a disk can be found  that is entirely convective, however, again on the basis of Fig.~\ref{fig:consistency} it may be likely that many discs are convective at least in some regions. This being the case one can speculate on the structure of discs that are partially convective. If such discs were capable of dynamo action what would the magnetic field look like? Would its structure be determined by the convective regions, or the radiative regions? Would it be possible to address some of these issues, at least in part, by local, i.e shearing box simulations, or would any analysis require  full disk simulations?

Finally, we note that the main objective of the present paper was to describe a procedure, not really to apply it to any particular situation. The example we have given was meant to be illustrative.  Applications to specific discs require  much more careful discussions of the physical conditions of the plasma,  and in particular of the opacity laws--most likely involving table lookups. Such analyses will be undertaken elsewhere.

\section{Acknowledgment}
This work was supported in part  by the National Science Foundation 
sponsored Center for Magnetic Self Organization at the University of Chicago. 
We acknowledge that the results in this paper have been achieved using the 
PRACE Research Infrastructure resource FERMI based in Italy at  the Cineca Supercomputing Center. The authors would like to thank the anonymous referee for many useful comments.


\begin{thebibliography}{14}
\expandafter\ifx\csname natexlab\endcsname\relax\def\natexlab#1{#1}\fi

\bibitem[{{Bodo} {et~al.}(2012){Bodo}, {Cattaneo}, {Mignone}, \&
  {Rossi}}]{Bodo12}
{Bodo}, G., {Cattaneo}, F., {Mignone}, A., \& {Rossi}, P. 2012, \apj, 761, 116

\bibitem[{{Bodo} {et~al.}(2013){Bodo}, {Cattaneo}, {Mignone}, \&
  {Rossi}}]{Bodo13}
---. 2013, \apjl, 771, L23

\bibitem[{{Bodo} {et~al.}(2015){Bodo}, {Cattaneo}, {Mignone}, \&
  {Rossi}}]{Bodo15}
---. 2015, \apj, 799, 20

\bibitem[{{Cabot}(1996)}]{Cabot96}
{Cabot}, W. 1996, \apj, 465, 874

\bibitem[{Hansen {et~al.}(2004)Hansen, Kawaler, \& Trimble}]{Hansen04}
Hansen, C., Kawaler, S., \& Trimble, V. 2004, Stellar Interiors: Physical
  Principles, Structure, and Evolution, Astronomy and Astrophysics Library
  (Springer New York)

\bibitem[{{Hawley} {et~al.}(1995){Hawley}, {Gammie}, \& {Balbus}}]{Hawley95}
{Hawley}, J.~F., {Gammie}, C.~F., \& {Balbus}, S.~A. 1995, \apj, 440, 742

\bibitem[{{Hayashi}(1961)}]{Hayashi61}
{Hayashi}, C. 1961, \pasj, 13, 450

\bibitem[{{Hirose}(2015)}]{Hirose15}
{Hirose}, S. 2015, \mnras, 448, 3105

\bibitem[{{Hirose} {et~al.}(2014){Hirose}, {Blaes}, {Krolik}, {Coleman}, \&
  {Sano}}]{Hirose14}
{Hirose}, S., {Blaes}, O., {Krolik}, J.~H., {Coleman}, M.~S.~B., \& {Sano}, T.
  2014, \apj, 787, 1

\bibitem[{Ponzo(2015)}]{Ponzo15}
Ponzo, F. 2015, Master's thesis, University of Torino

\bibitem[{Prialnik(2011)}]{Prialnik11}
Prialnik, D. 2011, An Introduction to the Theory of Stellar Structure and
  Evolution (Cambridge University Press)

\bibitem[{{Pringle}(1981)}]{Pringle81}
{Pringle}, J.~E. 1981, \araa, 19, 137

\bibitem[{{Shakura} \& {Sunyaev}(1973)}]{SS73}
{Shakura}, N.~I., \& {Sunyaev}, R.~A. 1973, \aap, 24, 337

\bibitem[{{Stone} \& {Balbus}(1996)}]{Stone96}
{Stone}, J.~M., \& {Balbus}, S.~A. 1996, \apj, 464, 364

\end{thebibliography}
\end{document}